\documentclass[12pt]{article}
\usepackage{float}
\usepackage{epsfig}

\newcommand{\be}{\begin{equation}}
\newcommand{\ee}{\end{equation}}
\newcommand{\bea}{\begin{eqnarray}}
\newcommand{\eea}{\end{eqnarray}}

\newcommand{\rme}{\mathrm{e}}
\newcommand{\rmi}{\mathrm{i}}

\newcommand{\rmc}{\mathrm{c}}
\newcommand{\rmH}{\mathrm{H}}
\newcommand{\rmL}{\mathrm{L}}

\begin{document}

\begin{center}
{\Large \bf The Nonperiodic Anyon Model and the Fractional
Quantum Hall Effect}\\[0.5cm]

{\large \bf Stefan Mashkevich}\footnote{mash@mashke.org}\\[0.1cm]
Schr\"odinger, 120 West 45th St., New York, NY 10036, USA\\[0.2cm]
{\large \bf St\'ephane Ouvry}\footnote{ouvry@lptms.u-psud.fr}\\[0.1cm]
Laboratoire de Physique Th\'eorique et Mod\`eles
Statistiques\footnote{{\it Unit\'e Mixte de
CNRS --- Universit\'e Paris Sud, UMR 8626}}\\
B\^at.~100, Universit\'e Paris-Sud, 91405 Orsay, France
\\
\end{center}

\vskip 0.5cm
\centerline{\large \bf Abstract}
\vskip 0.2cm

The lowest-Landau-level anyon model
becomes nonperiodic in the statistics parameter when
the finite size of the attached flux tubes is taken into account.
The finite-size effects cause the inverse proportional relation
between the critical filling factor and the statistics parameter to
be nonperiodically continued in the screening regime, where the fluxes are anti-parallel
to the external magnetic field --- at critical filling,
the external magnetic field is entirely screened
by the mean magnetic field associated with the flux tubes.
A clustering argument is proposed to
select particular values of the statistics parameter.
In this way, IQHE and FQHE fillings are obtained
in terms of gapped nondegenerate LLL-anyonic
wave functions.  Jain's series are reproduced without the need to
populate higher Landau levels.  New FQHE series are proposed,
like, in particular, the particle-hole complementary series of the Laughlin one.
For fast-rotating Bose-Einstein condensates, a corresponding clustering argument
yields particular fractional filling series.

\section{Introduction}

Very soon upon the discovery of the quantum Hall effect (QHE),
it was realized that it has to do with the very basic features
of quantum statistics in two dimensions.
Recall that the dimensionless Hall conductance
$2\pi\sigma_\rmH/e^2$ ($\hbar = 1$)
of a two-dimensional electronic sample in a strong magnetic field $B$
takes on values which are either
integers (IQHE) \cite{vK1980} or simple fractions (FQHE) \cite{Tsui1982}.
It can be shown \cite{PG1990} to be equal to the electronic filling factor 
\be \nu={\rho\over \rho_\rmL} \;,\ee
i.e., the ratio of the electron density $\rho$
(typically $\sim 10^{11}\: \mathrm{cm}^{-2}$;
the electrons are assumed to be fully spin polarized)
to the Landau level density (the degeneracy of a Landau level per unit area)
\be
\rho_\rmL = \frac{B}{\phi_0} \;,
\label{rhol}
\ee
where $\phi_0 = 2\pi/e$ is the flux quantum
($eB>0$ has been assumed without loss of generality).

The IQHE is explained in terms of free electrons, which in their ground state
fill an integer number $\nu$ of Landau levels, the ``excessive''
electrons being localized on impurities;
hence the conductance remains constant over a finite range of magnetic field strength,
as long as the Fermi level lies within a mobility gap.
The first excited state is separated by the cyclotron gap.

For the FQHE, where $\nu$ is a fraction, interaction between electrons is crucial.
Multielectron states corresponding to the Laughlin filling factors
\be
\nu = \frac{1}{2m+1}
\label{nul}
\ee
 ($m$ is an integer) are described by the antisymmetric Laughlin
ground state wave function \cite{Laugh1983}
\be\label{function}
\psi = \prod_{i<j} (z_i-z_j)^{2m+1} \exp\left(-{1\over 2}\omega_\rmc
\sum_i z_i\bar z_i\right) \;
\label{psil}
\ee
($\omega_\rmc = eB/2$ is half the cyclotron frequency).
Here all the particles are in the lowest Landau level (LLL), since the
wave function, less the exponential damping factor, is purely analytic
[the single-particle LLL eigenstates are $z^l \exp(-{1\over 2}\omega_{\rm
c} z\bar z), \; l \ge 0$].
Electrons in this state condense into a so-called incompressible
quantum fluid; again, there is a gap, which is crucial for the effect.

Explaining other fractions, however, is more challenging.
The concept of ``hierarchies'', brought forward by Haldane \cite{Haldane1983},
dwells on the fact that quasihole/quasiparticle excitations in the Laughlin model
obey fractional statistics \cite{anyon} with statistics parameter $\pm 1/(2m+1)$.
Assume that those excitations themselves condense into a Laughlin-like
state, but now, taking into account their statistics, the Slater determinant 
has to be raised to the power $\pm 1/(2m+1) + 2p$, where $p$ is another integer.
The resulting filling factor for the original electrons is
$\nu = 1/(2m+1 \pm 1/2p)$.
Building up excitations upon the gas of excitations and repeating the argument
leads to the general expression
\be\label{lolo}
\nu = \frac{1}{2m+1 \pm 
\displaystyle \frac{1}{2p_1 \pm \displaystyle \frac{1}{2p_2 \pm \displaystyle
\frac{1}{2p_3 \pm \cdots}}}} \;
\ee
(the fraction can be cut off after any $p_i$).
A few particular cases are: $\frac27$, $\frac25$, $\frac23$, $\frac45$
at the first level of the hierarchy, $\frac35$ at the second level.
However, some experimentally observed fractions can only be obtained
at quite high levels of the hierarchy, when there are more quasiparticles
than particles, raising concern about the validity of the concept of quasiparticles
in those cases.

A model which claims to be devoid of this shortcoming
is Jain's ``composite fermions'' \cite{Jain}.
It assumes that the combined effect of the strong magnetic field and of the
interaction between electrons, in the LLL, 
is to attach $2m$ flux quanta to each electron,
not changing their statistics,
but partially screening the external magnetic field (in a mean-field sense).
The FQHE is then understood as the IQHE with filling $\nu^* = p$ in
the screened field (projection on the LLL yields Jain's wave functions).
Relating the filling factor back to the original field gives the Jain series ($m\ge 0$)
\be
\nu={1\over 2m + \frac{1}{p}} \quad {\rm or}\quad \nu={1\over 2(m+1) - \frac{1}{p}} \;.
\label{nuj}
\ee
This obviously reduces to the IQHE for $m=0$
and to Laughlin's fractions for $p=1$.
In the general case, though, the LLL projection of the wave functions
is difficult to perform and leads to cumbersome expressions.

In this paper, we propose an alternative scenario of the QHE,
where it is assumed that an arbitrary fraction
of the flux quantum gets attached to each electron.
This leads to an effective change of the statistics of the electrons,
turning them into anyons \cite{anyon}.
By convention, those can be equivalently described as bosons
to which a flux $\alpha\phi_0$ is attached,
where $\alpha$ is the statistics parameter;
even/odd values of $\alpha$ correspond to Bose/Fermi statistics.
Interchanging two such objects generates an Aharonov-Bohm phase $\exp(-\rmi\pi\alpha)$.
It happens that anyons on the LLL constitute an exactly solvable two-dimensional model
with fractional statistics. It has been shown \cite{LLLanyons}
that the critical filling factor, which is the
maximal number of particles that can occupy a one-body quantum state
in the LLL, is simply
\be\label{fillalpha}
\nu=-{1\over\alpha} \;.
\ee
At this filling, where the external magnetic field is entirely screened
by the attached fluxes (from a mean-field point of view),
the multiparticle ground state becomes nondegenerate:
\be
\psi= \prod_{i<j} (z_i-z_j)^{-\alpha} \exp\left(-{1\over 2}\omega_\rmc
\sum_i z_i\bar z_i\right) \;,
\label{wfalpha}
\ee
and the pressure diverges.

One immediately notes the similarity between Eqs.~(\ref{fillalpha})--(\ref{wfalpha})
and (\ref{nul})--(\ref{psil}),
respectively, which is apparently not accidental --- except that in the case at hand,
$\alpha$ can be fractional,
thus eliminating the need for hierarchies or higher Landau levels.

However, two questions have to be answered in order
to render this approach consistent with the experimental QHE situation:
\begin{itemize}
\item[(i)]
Since the anyonic phase factor is periodic in $\alpha$ with period 2,
so are the anyon wave functions, spectra, and therefore the filling factors.
This means that Eqs.~(\ref{fillalpha})--(\ref{wfalpha}),
in the above form, are only valid for $\alpha \in [-2,0]$.
When extending them onto the interval $\alpha \in [-2(m+1),-2m]$,
one has to replace $\alpha$ with $\alpha+2m$.
Hence, only values of $\nu \in [1/2,\infty]$ are obtainable.
How can one eliminate this restriction?
\item[(ii)] 
How can preferred values of $\alpha$, corresponding to
the observable values of $\nu$, possibly be selected
from a continuous set (\ref{fillalpha})?
\end{itemize}

Our answers to these questions are:
\begin{itemize}
\item[(i)] Assume that the attached flux tubes have a finite size, which
breaks the periodicity in $\alpha$.
In a certain approximation, this will mean
an analytic continuation of the critical filling curve
valid on the interval $\alpha\in[-2,0]$
into the region $\alpha < -2$.
\item[(ii)] In the interval $[-2,0]$, use a cluster argument
that considers as preferred values of the statistics
parameter the ones for which a cluster of $p$ anyons
carries an odd number of flux quanta (Fermi-like),
i.e., $\alpha=-2+1/p$, $\alpha=-1/p$, or $\alpha=-1\pm1/p$
(in the latter case, $p$ has to be even).
Indeed, if the original integer (fermionic) statistics of the electrons has been traded off,
at the microscopic level, for an anyonic statistics,
one still would like to recover an effective Fermi-like description of the system,
at least in terms of finite clusters. 
\end{itemize}
It then follows from (i) and (ii) that 
in the interval $[-2(m+1),-2m]$, the clustering condition modifies as,
respectively, $\alpha = -2(m+1) + 1/p$,
$\alpha = -2m - 1/p$, for which the critical filling (\ref{fillalpha}) 
implies the Jain series (\ref{nuj}),
or $\alpha = -(2m+1) \pm 1/p$ with $p=2q$, which
predicts an FQHE series outside the Laughlin-Jain ones:
\be  \nu = {2q\over 2(2m+1)q\mp 1} \;.\ee

In the next section, we review the standard LLL-anyon model. In Sec.~3,
the finite-size argument is introduced, which renders the model nonperiodic,
and, in Sec.~4, thermodynamic aspects of the model are considered.
In Sec.~5, the clustering argument is proposed, and
similar considerations are made for fast-rotating Bose-Einstein
condensates \cite{joli}. Concluding remarks are presented in Sec.~6.

\section{A reminder: the LLL-anyon model} 
One starts with pointlike anyons with statistics parameter $\alpha\in[-2,0]$.
Note that the Bose limits $\alpha\to 0^-$ and $\alpha\to -2^+$
have no reason to be identical, since there is a strong perpendicular magnetic field.
\begin{figure}
\centerline{\epsfig{figure=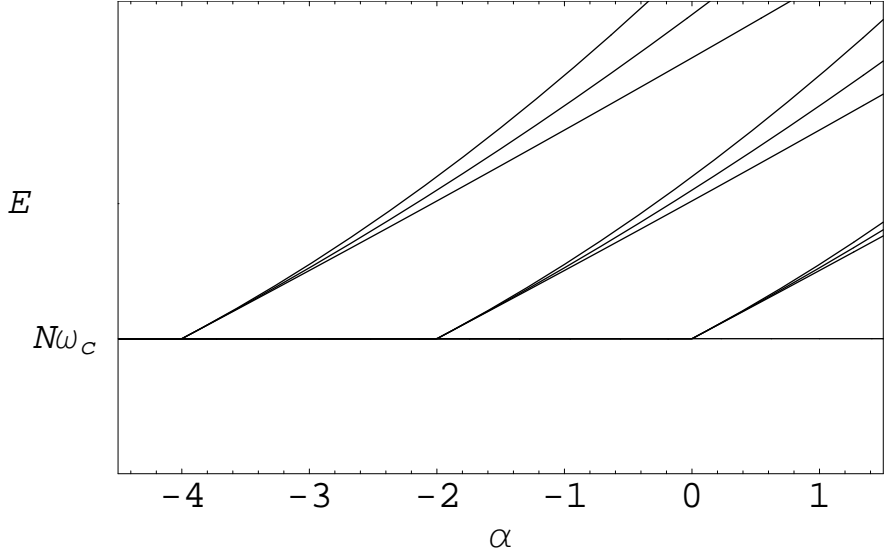,width=12cm}}
\caption{LLL anyonic eigenstates at $E_N=N\omega_{\rm c}$,
and the unknown eigenstates with a nonlinear dependence $E(\alpha)$
joining the ground state at the bosonic values of $\alpha$.}
\label{fig1}
\end{figure}

When $\alpha \in [-1,0]$, the LLL $N$-anyon eigenstates, interpolating
between the LLL-Bose $(\alpha=0)$ and LLL-Fermi ($\alpha=-1$) ones,
are known exactly:
\be
\label{excitation}\psi =
\prod_{i<j} (z_i-z_j)^{-\alpha}\prod_i z_i^{l_i}
\exp\left(-{1\over 2}\omega_{\rm c} \sum_i z_i\bar z_i\right) \;,
\quad 0\le l_1\le l_2 \le \ldots \;.
\ee
For a strictly infinite system, these are infinitely degenerate, with energy
$E=N\omega_\rmc$, and there is a gap above the LLL spectrum.
At each bosonic point $\alpha = 0,-2,-4,...$, some eigenstates, whose
wave functions are not known exactly, merge into the ground state (Fig.~\ref{fig1}).
However, as $|\alpha|$ increases, so do
the absolute values of the kinetic angular momentum of all states;
if there is a border, wave functions get ``pushed out'' beyond it, and energies increase.
This compensates for the states merging into the ground
state, and the total number of states in the LLL (averaged over an interval of
$\alpha$ of length 2) remains constant.
Thus, the physical quantities are periodic in $\alpha$ with period 2 ---
as they should be.

One obtains \cite{LLLanyons} the grand partition function
\be \label{gp}\ln Z={\rho_LV}\ln y \;,\ee
where $V$ is the area of the two-dimensional sample, and $y$
is related to the fugacity $z$ as
\be \label{y} y-z\rme^{-\beta\omega_\rmc}y^{1+\alpha}=1 \;;\ee
hence the equation of state 
\be\label{state}
\beta P = \rho_{\rm L} \ln \left(1+{\nu\over 1+\alpha\nu}\right) \;.
\ee
At $\nu \ll 1$ this turns into the equation of state of an ideal gas,
whereas at the critical filling $\nu=-1/\alpha$ 
the pressure diverges and the nondegenerate gapped LLL ground state is
given by Eq.~(\ref{excitation}) with all $l_i = 0$, i.e., Eq.~(\ref{wfalpha}).
Note that the total angular momentum
of this nondegenerate state is related
to the critical filling (\ref{fillalpha}) by
\be\label{fillingbis}  L={N(N-1)\over 2\nu} \;.\ee

\begin{figure}
\centerline{\epsfig{figure=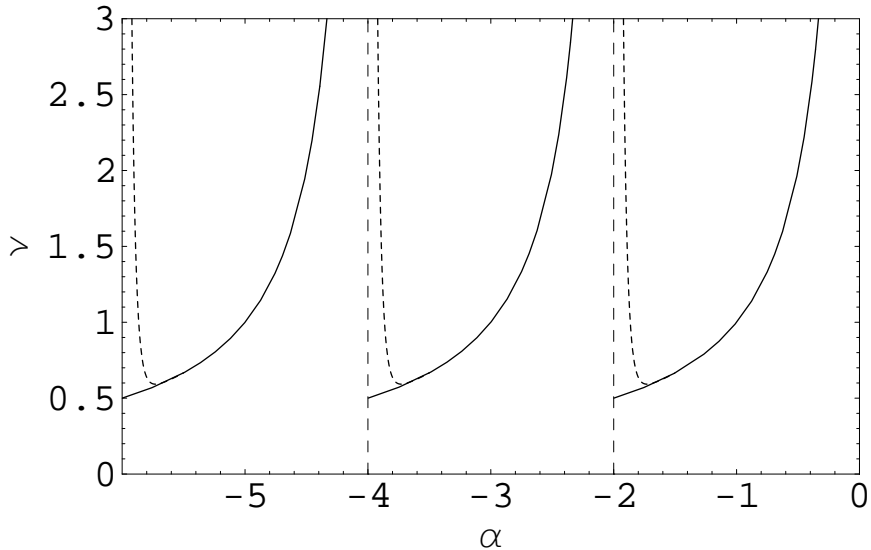,width=12cm}}
\caption{The critical filling curve continued periodically (solid curve). 
The jump at the Bose points would be eliminated (i.e., $1/\nu$
would be continuous) if the unknown eigenstates joining
the ground state at these points were taken into account (dashed curve).}
\label{fig2}
\end{figure}

Further, when $\alpha\in[-2,-1]$, it has been shown \cite{MO2003} that
Eqs.~(\ref{excitation})--(\ref{fillingbis}) obtained on the
interval $\alpha\in[-1,0]$ remain valid, up to the Bose
point   $\alpha=-2$ where the gap disappears due
to the unknown eigenstates joining the LLL at $\alpha\to-2^+$ (Fig.~\ref{fig1}).
More precisely, the gap is proportional to $(\alpha+2)\omega_{\rm c}$,
which means that the unknown states manifest themselves only in a narrow region
($\sim 1/\omega_{\rm c}$ wide) to the right of the Bose point.
In that region, the thermodynamics changes quickly but continuously, the critical
filling increasing from slightly above $1/2$ to $\infty$ (Fig.~\ref{fig2}).
If, however, one takes the $\omega_{\rm c} \to\infty$ limit first,
the states in question can be disregarded, but the
thermodynamics will become discontinuous at the Bose point.
Following this approach, the equation of state is still given by Eq.~(\ref{state}),
the LLL critical filling by Eq.~(\ref{fillalpha}) 
and the corresponding nondegenerate LLL ground state and LLL
excitations are Eqs.~(\ref{wfalpha}) and (\ref{excitation}), respectively.

Up to this point, apart from the fermionic critical filling $\nu=1$ at $\alpha=-1$,
i.e., the standard situation of an LLL completely filled with Fermi particles,
only one fraction is singled out: $\nu=1/2$, corresponding to $\alpha\to -2^+$.
Otherwise, the fillings form a continuous set in the interval $[1/2,\infty]$.
Clearly, some arguments have to be developed
in order to extend the critical filling interval
to $[0,\infty]$ and to arrive at observable fractions.

\section{Nonperiodic LLL-anyon model} 

For $\alpha\in[-2,0]$, the physics of the LLL-anyon model is (surprisingly) well
described by a mean-field picture, whereby one replaces the anyon fluxes
with a constant magnetic field $\rho \alpha \phi_0$ (preserving the total flux).
The critical filling, in this approach, corresponds to a complete
screening\footnote{Note that in Ref.~\cite{MO2003}, 
this regime in the interval $\alpha\in[-2,-1]$\ is
equivalently described as anti-screening,
by considering $\alpha\in[0,1]$ instead.
One then has $B+\rho(\alpha-2)\phi_0=0$, i.e., the external
magnetic field and the mean field carried by the vortices add up to
$2\rho\phi_0$.}
of the external magnetic field by this mean field.
Indeed, taking into account Eq.~(\ref{fillalpha}), the critical filling condition  implies
\be\label{screening} B + \rho\alpha\phi_0=0 \;.\ee

As $\rho$ increases, the total magnetic field decreases, due to screening,
and so does the effective Landau level density (the number of states
available for subsequent particles). At the critical filling, it drops
to zero, meaning no more particles can be added.

In order for the model being advocated here to be relevant for the FQHE,
this picture should be valid for any $\alpha$.
However, with pointlike Aharonov-Bohm fluxes, the mean-field picture
(\ref{screening}) is only valid for $\alpha\in[-2,0]$,
because of the periodicity in $\alpha$.

If the fluxes have a finite size, there is no more periodicity.
This is physically sound, since pointlike fluxes are
an idealization to begin with; being effectively induced by the
interaction and the external magnetic field, they are expected
to have a size of the order of the magnetic length.
The Aharonov-Bohm periodicity argument is 
no longer pertinent as soon as the size becomes finite,
and the mean-field picture should be more accurate in this case.
We will now argue that the eigenstates
which, for pointlike fluxes, join the LLL at $\alpha = -2, -4, \ldots$
(leading to periodicity), will no longer do so for finite-size fluxes ---
resulting in the validity of the wave functions (\ref{wfalpha})
and filling (\ref{fillalpha}) beyond $\alpha = -2$.
Indeed, at the point where a state joins the LLL,
so its energy is a nonanalytic function of $\alpha$,
its wave function does not vanish at zero interparticle distance
(like in a state of two bosons with zero relative angular momentum).
On the other hand, with a finite size, there appears an effective short-range repulsion,
due to the nonzero magnetic field in the vicinity of a particle.
This repulsion will lift the energies of the states
whose wave functions do not vanish when particles collide,
while (almost) not affecting states whose wave functions do vanish.
The states joining the LLL at $\alpha = 0$ will be unaffected,
since there is no magnetic field in them.
The wave functions are no longer exact,
but are expected to be a good approximation to the exact ones
(just like the original Laughlin functions) at least on some interval of $\alpha$.

\begin{figure}
\centerline{\epsfig{figure=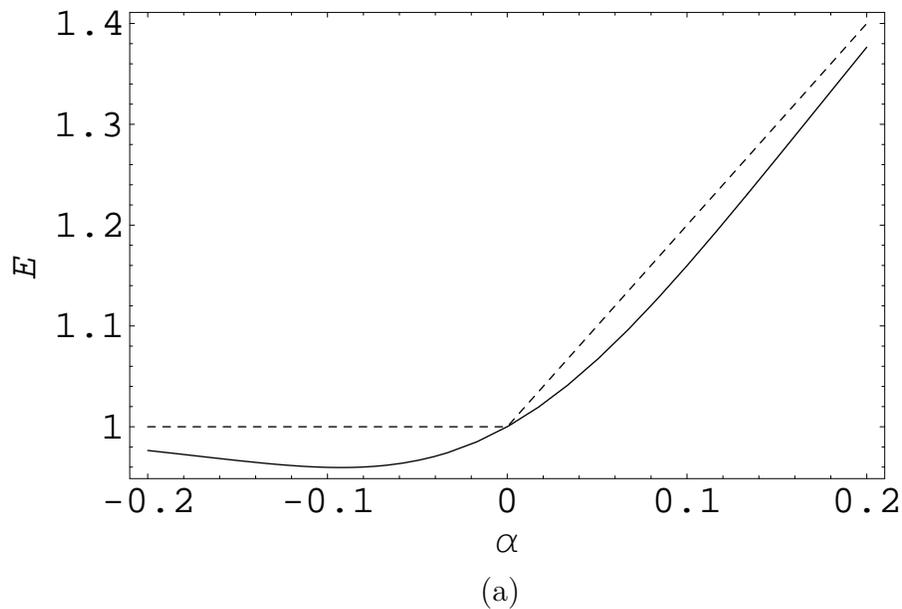,width=12cm}}
\centerline{\hskip 3em (a)}
\vskip 4em
\centerline{\epsfig{figure=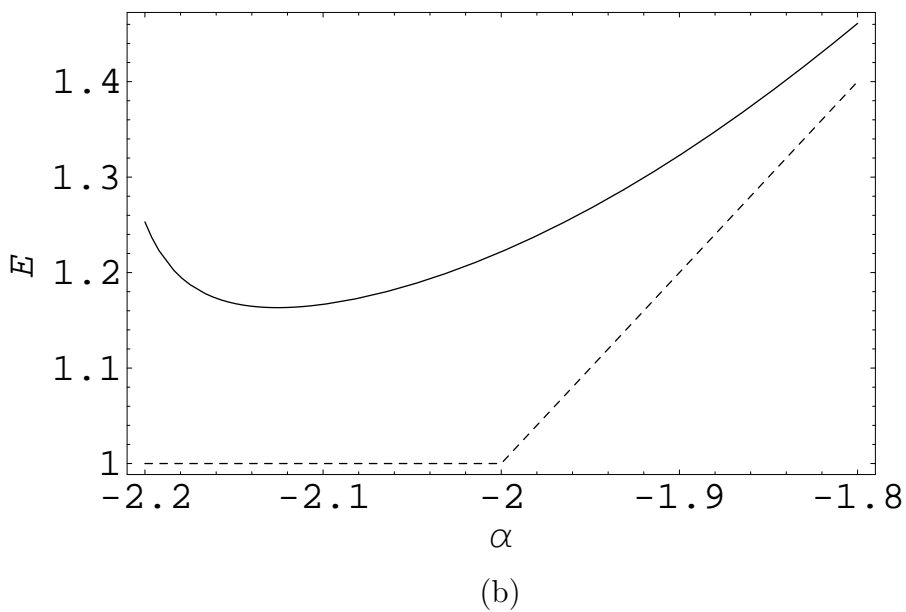,width=12cm}}
\centerline{\hskip 3em (b)}
\caption{The two-anyon states, $l=0$ (a) and $l=2$ (b),
with a finite flux tube radius $R=0.01$ (solid curves),
compared to the corresponding pointlike anyon states (dashed lines); $\omega_\rmc = 1$.
The former still joins the LLL at $\alpha = 0$, the latter
no longer does so at $\alpha = -2$, being lifted
due to the effective repulsion from the flux tube.}
\label{fig3}
\end{figure}

By way of supporting this argument quantitatively, consider the 
problem of two bosons with a tube of radius $R$ carrying a flux $\alpha\phi_0$
attached to each \cite{SM96}.
The center-of-mass motion detaches as usual, and the problem of relative
motion is that of a single particle in the external magnetic field
plus the tube field.
Outside the tube, $r>R$,
the radial part of the wave function with energy $E$, regular at infinity, is
\be
\psi_{\rm out} = r^{|l+\alpha|}
M\left(\frac{|l+\alpha| - (l+\alpha) + 1 - E/\omega_\rmc}{2}, \: 
|l+\alpha|+1, \: \omega_\rmc r^2 \right)
\rme^{-\frac{\omega_\rmc r^2}{2}} \;,
\ee
where $l$ is even, but the kinetic
angular momentum is $l+\alpha$, due to the flux of the tube.
Let the magnetic field inside the tube be uniform,
$B_\alpha = \alpha\phi_0/(\pi R^2) = 2\alpha/(eR^2)$;
then the total field, $B+B_\alpha$, is also uniform,
and the wave function for $r<R$, regular in the origin, is
\be
\psi_{\rm in} = r^{|l|} U\left(\frac{|l| - l + 1 - E/|\omega'_\rmc|}{2},
\: |l|+1, \: |\omega'_\rmc| r^2 \right)
\rme^{-\frac{|\omega'_\rmc| r^2}{2}} \;
\ee
with $\omega'_\rmc = \omega_\rmc + \alpha/R^2$
($M$ and $U$ are the confluent hypergeometric functions
of the first and second kind, respectively).
Equating the logarithmic derivative at $r=R$ yields the energy levels.
For $R = 0$, the condition is simply that $\psi_{\rm out}$ be regular at the origin,
implying $[|l+\alpha| - (l+\alpha) + 1 - E/\omega_\rmc]/2 = -n$.
Hence the familiar LLL ($n=0$) spectrum,
\be
E = [|l+\alpha| - (l+\alpha) + 1]\omega_\rmc \;
\ee
--- at each of $\alpha = -2, -4, \ldots$, a state joins the LLL 
(the relative angular momentum of that state vanishes at that point).
However, with a finite $R$, finding the spectrum numerically,
one sees that the $l=0$ state is practically
unaffected, whereas the $l=2$ state no longer joins the LLL
at $\alpha = -2$ (Fig.~\ref{fig3}).

{}From the above consideration one concludes that equations
(\ref{excitation})--(\ref{fillingbis}), valid on the interval $\alpha\in [-2,0]$,  
can be analytically continued beyond $\alpha = -2$
provided that the finite-size effects are taken into account.
At distances much bigger than the flux tube radius, the particles
still behave like anyons with statistics parameter $\alpha$;
however, there is no longer a complete Bose/Fermi interpolation
when $\alpha$ varies from $-2m$ to $-2m-1$.

\section{Thermodynamics}

{}From the thermodynamic point of view, the analytic continuation
beyond $\alpha\in[-2,0]$ also makes sense.
In the LLL-anyon model,  one infers
from Eqs.~(\ref{gp})--(\ref{y}) that the LLL $N$-anyon partition function $Z_N$,
i.e. the coefficient at order $N$ of the expansion of $Z$
in powers of the fugacity $z$, is, for $\alpha\in[-2,0]$,
\be\label{ZN}
Z=1+\rho_L V \sum_{N=1}^\infty Z_N z^N \;,
\ee
where
\be\label{comb} Z_N = \prod_{k=2}^N
\left(1+{\rho_L V-1+\alpha N\over k}\right)\rme^{-N\beta\omega_\rmc}\ee
($Z_1 = 1$).
Since Eqs.~(\ref{ZN})--(\ref{comb}) follow directly
from the well-defined microscopic LLL-anyon quantum-mechanical model,
$Z_N$ has to be a well-defined positive quantity for all $N$
--- which it is, since the LLL degeneracy $\rho_L V$
is proportional to the surface of the system.

In the nonperiodic generalization of the model at hand,
$Z_N$ remains a positive quantity when $\alpha$ is analytically continued
beyond $\alpha = -2$, provided that $\rho_LV>-\alpha N$, i.e., $\nu<-1/\alpha$,
which is indeed enforced by Eq.~(\ref{fillalpha}).
Therefore, the finite-size flux tube model
is also well defined at the thermodynamic level.

One can push this analysis further and give a combinatorial meaning to $Z_N$:
leaving aside the thermal factor $\rme^{-N\beta\omega_\rmc}$,
it is well known \cite{Poly} that $Z_N$ has a simple combinatorial  interpretation,
at least for an integer $-\alpha$.  It counts the number of ways to put $N$ particles
into $\rho_L V$ degenerate quantum states arranged  on a circle, with the condition that there are
at least $-\alpha-1$ empty states between two occupied states. At the critical filling
$\nu=-1/\alpha$, this condition narrows down to there being precisely $-\alpha-1$
empty states between two occupied ones.
(This is the generalization of the Pauli principle known as the Haldane
exclusion rule \cite{Haldane}: adding a particle to the system makes
$-\alpha$ states unavailable for subsequent particles.)
The quantum states here are labelled
by the orbital angular momentum $l \ge 0$, meaning,
for  example, for the $-\alpha=2m+1$ Laughlin states with $m=1$,
an angular momentum occupation of the type
\be 100100100100100...\ee
Equivalently, in the Bose case $\alpha=-2$ at filling $1/2$, one would have
 \be 1010101010...\ee
For filling $2/5$, corresponding to $\alpha=-5/2$,
one infers from Eq.~(\ref{comb}) that the orbital momentum occupation
should be on average of the type  
\be 1001010010100101001010010...\ee
This can be generalized to any fractional filling governed by (\ref{comb}).
Similar conclusions for fractional fillings have been obtained in a different context
(one-dimensional thin torus model approach to the QHE) in Ref.~\cite{Kar}.

\section{Cluster argument} 

In the composite-fermion theory of the FQHE, the integer parameter $p$
can be reinterpreted in the LLL
\cite{KvKE99}  by a clustering argument ($p$ is the number of particles per cluster); in this
scheme, Jain's wave functions are obtained directly in the LLL (without going on
the $p$-th Landau level and then projecting on the LLL).

The clustering point of view can be easily understood in the 
LLL-anyon context: in the interval $[-2,0]$, if one sets 
\be\label{11} \alpha=-1/p\quad{\rm or}\quad \alpha=-2+1/p \;, \ee
one obtains for the critical fillings
\be
\quad \nu={p}  \quad {\rm or} \quad  \nu={1\over 2-1/p} \;,
\ee
the former being the IQHE series (as already proposed in Ref.~\cite{LLLanyons}),
the latter a special case of the Jain series (\ref{nuj}).
In both these cases, the total magnetic flux attached to a cluster
(in the units of the flux quantum) is odd, i.e., Fermi-like.
We claim, therefore, that these are among the favored values of $\alpha$.
Note, however, that the Bose points $\alpha=0,-2$
are excluded since the size of the cluster would then be infinite.

Similarly a cluster of $p$ anyons
with 
\be\label{22}
\alpha=-1\pm 1/p\ee
is also Fermi-like if $p=2q$ is even, leading to
\be
\nu={1\over 1- 1/(2q)} \quad {\rm or}\quad \nu={1\over 1+ 1/(2q)} \;,
\ee
the former being a $\nu>1$ series and the latter [statistics $1+ 1/(2q)$,
filling $2q/(2q+1)$] corresponding to the particle-hole
complementary of the Laughlin wave function [filling $1-\nu$, where $\nu$ is given by
Eq.~(\ref{nul}) with $m\to q$].

Considering now the nonperiodic continuation of the LLL-anyon model,
one generalizes the cluster ansatz (\ref{11})--(\ref{22}) from the interval $\alpha\in[-2,0]$
onto the interval $\alpha\in[-2(m+1),-2m]$, to get
\be\label{jain}\alpha=-1/p \quad \to \quad\alpha=-2m-1/p \quad {\rm thus}\quad
\nu={1\over 2m +{1/p}}\;,\ee
\be\label{jainbis}\alpha=-2+1/p \quad \to  \quad \alpha=-2(m+1)+1/p \quad
{\rm thus}\quad \nu={1\over 2(m+1) -{1/p}}\;,\ee
i.e., the usual Jain's series (\ref{nuj}), and
\be\label{jainter}\alpha=-1\pm 1/(2q) \quad \to
\quad\alpha=-(2m+1)\pm 1/(2q) \quad {\rm thus}\quad
\nu={1\over 2m +1\mp{1/(2q)}}\;.\ee
Note that the latter series, $\nu={2q\over 2(2m+1)q \mp 1}$, can be recast as
twice some fractions in the Jain series (\ref{jain})--(\ref{jainbis}).
In particular, for $m=1$ and $q=2$, the filling $\frac{4}{11}$ (twice 
$\frac{2}{11}$) has been observed experimentally.

Finally, the corresponding wave functions are
given by Eq.~(\ref{wfalpha}) with the appropriate
value of $\alpha$: for the series (\ref{jain})
one has the nondegenerate wave function
\be\label{lolobis}\psi=\prod_{i<j} (z_i-z_j)^{2m+{1\over p}}\exp\left(-{1\over 2}\omega_{\rm c} 
\sum_i z_i\bar z_i\right)\ee
with statistics $2m+1/p$
(see an argument in this direction in Ref.~\cite{CJW05}),
and similarly for the series in (\ref{jainbis}),
with statistics $2(m+1)-1/p$. 
Note that for $p=1$, both these LLL-anyon wave functions
coincide with the Laughlin functions (\ref{lolo}) at filling $1/(2m+1)$.
The series (\ref{jainter}) yields the nondegenerate wave function 
\be\label{loloquar}\psi=\prod_{i<j} (z_i-z_j)^{2m+1\mp {1\over 2q}}\exp\left(-{1\over 2}\omega_{\rm c} 
\sum_i z_i\bar z_i\right)\ee
with statistics $2m+1\mp 1/(2q)$.
[Note that the particle-hole
complementary of the Laughlin wave function, Eq.~(\ref{psil}) with $m\to q$,
is Eq.~(\ref{loloquar}) with $m=0$ and the plus sign.]

\begin{figure}
\centerline{\epsfig{figure=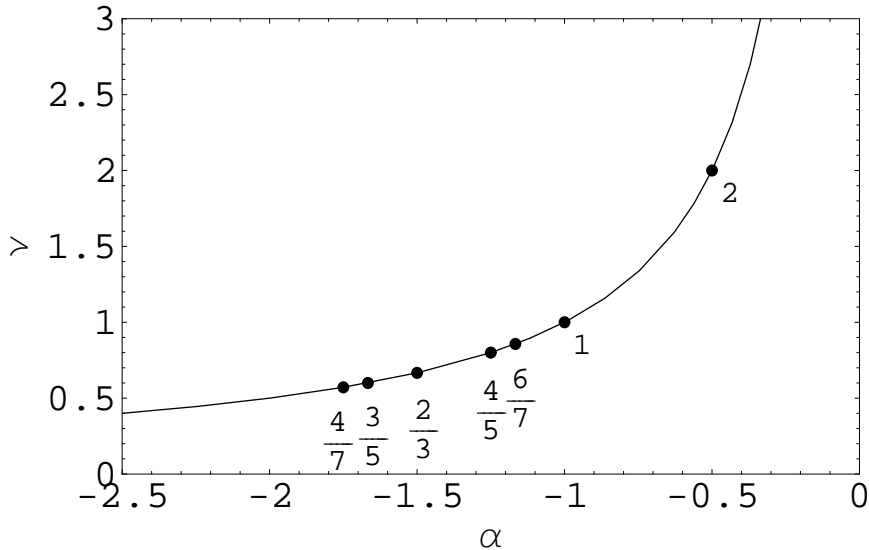,width=12cm}}
\caption{The critical filling curve for the nonperiodic LLL anyon model
on the negative $\alpha$ axis. The dots indicate some of the cluster ansatz 
induced rational values of $\nu$.}
\label{fig4}
\end{figure}

In this scheme, the IQHE and FQHE appear on the same footing, both described
by LLL-anyonic nondegenerate ground-state wave functions.
See Fig.~\ref{fig4} for an illustration of the various fillings obtained.
As already noted, the fillings $1/(2m)$ would require  infinite
clusters and are thus  excluded. At filling $p$, on the other hand,
leaving aside the anyonic phase, the short-distance behavior from
Eq.~(\ref{lolobis}) with $m=0$ is ${|z_i-z_j|}^{1/p}$.

Up to now one has considered clusters whose total flux is an odd number
of flux quanta. Having now in mind fast-rotating Bose-Einstein condensates,
one can also look at what particular fractional fillings
are obtained for clusters with an even total flux.
For $\alpha\in[-2,0]$, we are looking at a cluster of $p$ anyons
with $\alpha=-1 \pm 1/p$ and $p$ odd;
the corresponding critical fillings are
\be
 \nu={1\over 1\mp 1/(2q+1)} \;.
\ee
As before,
the analytic continuation beyond $\alpha = -2$ lets one generalize to
$\alpha\in[-2(m+1),-2m]$ with the special values $\alpha=-(2m+1)\pm 1/(2q+1)$
and corresponding fillings
\be
\nu={1\over 2m+1\mp 1/(2q+1)} \;.
\ee
This series corresponds to the nondegenerate wave function 
\be\psi=\prod_{i<j} (z_i-z_j)^{2m+1\mp {1\over 2q+1}}\exp\left(-{1\over 2}\omega_{\rm c} 
\sum_i z_i\bar z_i\right)\ee
with statistics $2m+1\mp 1/(2q+1)$.

It is interesting to note that fillings with
even denominators ($\nu = \frac{3}{8}$ and $\frac{3}{10}$, corresponding to $q=m=1$)
have been observed in recent FQHE experiments \cite{Pan}.

\section{Conclusion}

We have argued that the model of lowest-Landau-level anyons,
i.e., particles to which the combined effect of an external magnetic field
and Coulomb interactions effectively attaches fractional magnetic flux tubes,
loses its periodicity in the anyon statistics parameter
if the finite size of the flux tubes is taken into account.
This is due to a classical short-range interaction force caused by
the magnetic field inside the tube.
In this case 
the critical filling stays inversely proportional to the statistics parameter
beyond the natural interval $\alpha \in[-2,0]$.
The corresponding ground-state LLL anyon wave functions have the same structure
as the Laughlin functions, except that the Slater determinant is raised to
a fractional power.
These describe incompressible states of the anyon gas, and thereby explain
the integer and fractional quantum Hall effects on the same footing.

Selecting  statistics parameters  yielding simple rational
values of the critical filling
can be achieved via a clustering condition, whereby
the total flux of a finite cluster of anyons is an odd integer number of flux quanta.
This being taken into account, a two-parametric family of quantum Hall fractions
arises, which includes the Jain series as well as
other odd-denominator fractions (e.g., $\frac{4}{11}$).
Also, a complementary  Laughlin wave function, corresponding to
particle-hole symmetry [critical filling $1 - 1/(2m+1)$], arises naturally.
It is also notable that an alternative clustering condition,
calling for an even total flux of a cluster,
yields some experimentally-observed fractions with even denominators.

There appears a new distance scale in the problem, the radius of the flux tube.
For the model to be valid, this scale  needs to be ``optimal''.
Specifically, if one chooses a radius which tends to zero, one should recover the original
pointlike-particle, periodic model: this means that the domain of $\alpha$
on which the analytic continuation beyond $\alpha \in[-2,0]$ holds will be small.
On the other hand, if the radius is too big, the anyonic character of the
model will be lost, and the Laughlin-like ground-state LLL wave functions
cannot be expected to be valid.
It remains an open question to fully master the origin of this new scale.

We expect the approximation introduced to be valid on
a sufficient interval beyond $\alpha \in[-2,0]$ to encompass all the fractions
observed experimentally (the smallest experimental fraction
is of the order of $\frac{1}{7}$).
By way of some support for the clustering argument,
note that a recent study has shown that in two-dimensional fast-rotating
Fermi and Bose Fermi systems restricted to the LLL, 
delta-function repulsion can cause formation of multiparticle clusters \cite{vort}.

One of us (S.O.) would like to thank Thierry Jolicoeur for numerous
conversations on the quantum Hall effect.


\begin{thebibliography}{99}

\bibitem{vK1980} K. von Klitzing, G. Dorda, M. Pepper, Phys. Rev. Lett. 45, 494 (1980).
\bibitem{Tsui1982} D.C. Tsui, H.L. Stormer, A.C. Gossard, Phys. Rev. Lett. 48, 1559 (1982). 
\bibitem{PG1990} R. Prange, S. Girvin, {\em The Quantum Hall Effect}, Springer-Verlag (1990).
\bibitem{Laugh1983} R.B. Laughlin, Phys. Rev. B23, 3383 (1983).
\bibitem{Haldane1983} F.D.M. Haldane, Phys. Rev. Lett. 51, 605 (1983);
B.I. Halperin, Phys. Rev. Lett. 52, 1583 (1984).
\bibitem{anyon}
J.M.~Leinaas, J.~Myrheim, Nuovo Cimento 37B, 1 (1977);
G.A.~Goldin, R.~Menikoff, D.H.~Sharp, J.~Math.~Phys. 21, 650 (1980),
J.~Math.~Phys. 22, 1664 (1981);
F.~Wilczek, Phys.~Rev.~Lett. 48, 1144 (1982),
Phys.~Rev.~Lett. 49, 957 (1982).
\bibitem{Jain} J.K. Jain, Phys. Rev. Lett. 63, 199 (1989);
Phys. Rev. B40, 8079 (1989); Phys. Rev. B41, 7653 (1990);
see also the critical review of M. I. Dyakonov, cond-mat/0209206. 
\bibitem{LLLanyons} A.~Dasni\`eres de Veigy, S.~Ouvry, Phys. Rev. Lett. 72, 600 (1994).
\bibitem{joli} For a numerical analysis of this problem see, for example,
N. Regnault, Th. Jolicoeur, Phys. Rev. Lett. 91, 030402 (2003), Phys. Rev. B 69, 235309 (2004).
\bibitem{MO2003} S.~Mashkevich, S.~Ouvry, Phys. Lett. A310, 85 (2003).
\bibitem{SM96} S.~Mashkevich, Phys. Rev. D54, 6537 (1996).
\bibitem{Poly} A. Polychronakos, Generalized Statistics in One Dimension,
in {\em Topological Aspects of Low-Dimensional Systems}, 
Les Houches Summer School, Ed. A. Comtet, T. Jolicoeur,
S. Ouvry and F. David, Springer (1998), pp.~417--469.
\bibitem{Haldane} F.D.M. Haldane, Phys. Rev. Lett. 67, 937 (1991).
\bibitem{Kar}  E.J. Bergholtz, A. Karlhede, cond-mat/0509434.
\bibitem{KvKE99} I.V.~Kukushkin, K.~v.~Klitzing, K.~Eberl, Phys. Rev. Lett. 82, 3665 (1999).
\bibitem{CJW05} O. Ciftja, G. Japaridze, X.Q. Wang, J. Phys. Condens. Matter 17, 2977 (2005).
\bibitem{Pan}  W. Pan et al, Phys. Rev. Lett. 90, 0168101 (2003).
\bibitem{vort} S.M. Reimann, M. Koskinen, Y. Yu, M. Manninen, Phys. Rev. A 74, 043603 (2006).

\end{thebibliography}
\end{document}